\documentclass[10pt,twocolumn,letterpaper]{article}

\usepackage{wacv}
\usepackage{times}
\usepackage{epsfig}
\usepackage{graphicx}
\usepackage{amsmath}
\usepackage{amssymb}
\usepackage{float}
\usepackage{stfloats}
\usepackage{xcolor}
\usepackage{booktabs}
\usepackage{subcaption} 
\usepackage[final]{pdfpages}
\usepackage[utf8]{inputenc}

\def\wacvPaperID{514} 

\wacvfinalcopy 
\ifwacvfinal
\def\assignedStartPage{9876} 
\fi

\ifwacvfinal
\usepackage[breaklinks=true,bookmarks=false]{hyperref}
\else
\usepackage[pagebackref=true,breaklinks=true,colorlinks,bookmarks=false]{hyperref}
\fi

\ifwacvfinal
\else
\fi
\begin{document}

\title{Saliency Driven Perceptual Image Compression}

\author{Yash Patel\textsuperscript{2}\thanks{The reserch was conduceted during Y. Patel's internship at AWS.}, Srikar Appalaraju\textsuperscript{1}, R. Manmatha\textsuperscript{1}\\
	\textsuperscript{1}Amazon Web Services, Palo Alto\\
	\textsuperscript{2}Visual Recognition Group, Czech Technical University in Prague\\
	{\tt\small patelyas@fel.cvut.cz, (srikara,manmatha)@amazon.com}
}

\maketitle

\begin{abstract}
This paper proposes a new end-to-end trainable model for lossy image compression, which includes several novel components. The method incorporates 1) an adequate perceptual similarity metric; 2) saliency in the images; 3) a hierarchical auto-regressive model. This paper demonstrates that the popularly used evaluations metrics such as MS-SSIM and PSNR are inadequate for judging the performance of image compression techniques as they do not align with the human perception of similarity. Alternatively, a new metric is proposed, which is learned on perceptual similarity data specific to image compression. The proposed compression model incorporates the salient regions and optimizes on the proposed perceptual similarity metric. The model not only generates images which are visually better but also gives superior performance for subsequent computer vision tasks such as object detection and segmentation when compared to existing engineered or learned compression techniques.

\end{abstract}

\section{Introduction}
\label{sec:introduction}

In the last two decades, the number of images captured and transmitted via the internet has grown exponentially \cite{forbes}. This surge has increased both data storage and transmission requirements. Compression of images can be lossless \cite{boutell1997png,skodras2001jpeg}, that is, the original image can be perfectly reconstructed. A better compression rate can be obtained by using lossy methods such as JPEG \cite{wallace1992jpeg}, JPEG-2000 \cite{skodras2001jpeg} and BPG \cite{bpg}. The objective of these lossy methods is to obtain higher compression by removing the information which is least noticeable to humans. Note that these traditional codecs are hand-crafted and are not learned from the data.

More recent methods focus on learning to compress images. While learned image compression from data using neural networks is not new \cite{munro1989image,jiang1999image,luttrell1988image}, there has recently been a resurgence of deep learning-based techniques for solving this problem \cite{balle2016end,balle2018variational,mentzer2018conditional,rippel2017real,lee2018context,patel2019deep,theis2017lossy}. These models are trained by jointly minimizing rate (storage or transmission requirement) and distortion (reconstruction quality), leading to a Rate-Distortion trade-off \cite{shannon1948mathematical}.

The contributions of this paper are three-fold:  
\begin{enumerate}
    \item A novel hierarchical auto-regressive architecture for image compression is proposed, which is based on an encoder-decoder setup.
    \item A learned perceptual similarity metric is proposed. The metric is realized by a fully convolutional network (FCN) which is trained on compression specific artefacts. Note that the metric is differentiable and it is used as a for training the image compression model.
    \item The proposed model accounts for salient regions. This is achieved in two ways: (a) more bits are allocated to the salient regions and (b) higher weight is given to their reconstruction.
    \end{enumerate}

The motivations and background for these contributions are subsequently discussed, along with the related work.

\vspace{-0.3cm}
\paragraph{Taxonomy of image compression models.} Deep learning models for image compression can be broadly categorized into three generative models: a) Variational Auto-Encoders \cite{kingma2013auto,balle2018variational,balle2016end,lee2018context}; b) Generative Adversarial Networks \cite{rippel2017real,agustsson2018generative} (GAN) \cite{goodfellow2014generative}; c) Auto-Regressive (AR) \cite{van2016conditional,mentzer2018conditional}. Variational auto-encoders (VAEs) and auto-regressive models operate by estimating the probability density explicitly. On the other hand, GANs have an implicit measure of the density \cite{goodfellow2014generative}. GANs are useful in very low bit-rate settings as they can learn to synthesize the images \cite{agustsson2018generative}. However, their superiority over AR and VAE is unclear for higher bit-rates. A key difference between VAE and AR is that the former approximates the density, whereas auto-regressive models such as Pixel-CNNs, Pixel-RNNs \cite{van2016conditional} have an explicit and tractable measure of density either in the pixel space or in the learned quantized space.

\vspace{-0.3cm}
\paragraph{Proposed hierarchical auto-regressive model.} A hierarchical auto-regressive model with two-stages is designed, which is realized by two 3D Pixel-CNN \cite{van2016conditional}. The Pixel-CNNs operate on the learned quantized representations. During training, the 3D Pixel-CNNs give an explicit measure of the entropy of the quantized representations, which via information theory directly relates to the bits required to store them after arithmetic coding \cite{marpe2003context}. Minimizing the estimated entropy leads to compression, while the reconstructed image should be as close to the original as possible. Note that auto-regressive models have been used before for compression \cite{mentzer2018conditional}, but, this is the first hierarchical auto-regressive model designed for the task.

\begin{figure*}
	\centering
	\includegraphics[width=\textwidth]{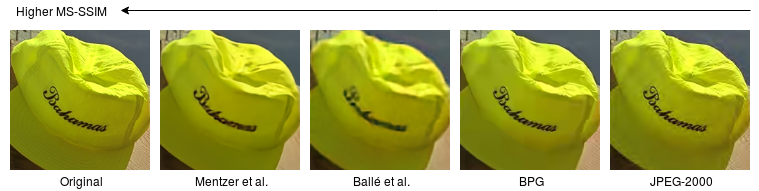}
	\caption{An example from the Kodak dataset \cite{kodak}. In order of MS-SSIM values: Mentzer ~\etal \cite{mentzer2018conditional} $>$ Ballé  ~\etal \cite{balle2016end} $>$ BPG \cite{bpg} $>$ JPEG-2000 \cite{skodras2001jpeg}. However, the order of performance based on $5$ human evaluations is: BPG \cite{bpg} $>$ Mentzer ~\etal \cite{mentzer2018conditional} $>$ JPEG-2000 \cite{skodras2001jpeg} $>$ Ballé  ~\etal \cite{balle2016end}. Visually the foreground and text in BPG is clearly better in quality.}
	\label{fig:kodak_samples}
	\vspace{-0.4cm}
\end{figure*}

\vspace{-0.3cm}
\paragraph{Existing evaluation metrics.} Lossy image compression models are compared by plotting the rate-distortion curve, where the rate is in bits-per-pixel vs the value of the distortion function. Common choices of the distortion function for compression models are MS-SSIM and PSNR \cite{balle2016end,balle2018variational,mentzer2018conditional,lee2018context}. Both of these metrics do not align well with human perception of similarity \cite{patel2019human}. As deep learned models directly optimize on these evaluation metrics, it is natural for them to have high MS-SSIM or PSNR scores when compared to the engineered methods such as JPEG-2000 \cite{skodras2001jpeg}, BPG \cite{bpg}. However, the resulting images often look worse to a human, \emph{i.e.}, the distortion in the images is higher although the metrics report otherwise. Figure \ref{fig:kodak_samples} shows four different techniques ranked in descending order of MS-SSIM values. The $2$nd and $3$rd images have many more artefacts than the last two images which imply that MS-SSIM is arguably not a good evaluation measure for compression. In Figure \ref{fig:kodak_samples} notice that the text is not as clear in the first two approaches.

The limitations of MS-SSIM and PSNR have been investigated in the past \cite{zhang18}. The super-resolution methods have started using a more sophisticated learned perceptual similarity metric \cite{blau2018perception,blau20182018} for evaluation, exactly for the mentioned reasons. Despite of the short comings, recent image compression literature continues to evaluate models using MS-SSIM and PSNR \cite{balle2016end,balle2018variational,mentzer2018conditional,lee2018context,patel2019deep,rippel2017real,theis2015generative,theis2017lossy}. Further, it has been observed that techniques trained on PSNR perform well when evaluated on PSNR, but poorly when evaluated using MS-SSIM (and vice-versa). This makes developing practical compression systems difficult. 

\vspace{-0.3cm}
\paragraph{Proposed evaluation metric.} This paper proposes a learned perceptual similarity metric for evaluating and training image compression models. This paper also studies the compression techniques using human evaluations. The study shows that human evaluation results correlate well with the proposed evaluation metric. Furthermore, the ranking of different compression methods, as judged by humans, correlates with the performance of off-the-shelf object detection and segmentation methods, which are trained on uncompressed images.

The analysis in this paper starts by evaluating the model proposed by Zhang ~\etal \cite{zhang18}. The model is trained on images with several different artefacts but the only compression artefacts are from JPEG. As Patel ~\etal \cite{patel2019deep} found, using the model \cite{zhang18} has limitations for compression. Therefore, we create a compression specific perceptual similarity dataset. The data include images generated from popular engineered \cite{skodras2001jpeg,wallace1992jpeg,bpg} and learned compression methods \cite{mentzer2018conditional,balle2016end,balle2018variational,patel2019deep}. The data consists of $6$ two alternative forced choices (2AFC) per comparison (see Section \ref{sec:percp_sim_metrices} for details). 

\vspace{-0.3cm}
\paragraph{Saliency matters for image compression.} While JPEG \cite{wallace1992jpeg} divides an image into uniform $8 \times 8$ blocks, BPG uses a hand-crafted metric to determine homogeneity and divides the more homogeneous regions into larger $64 \times 64$ blocks. In BPG, fewer bits are allocated to homogeneous regions and more bits are allocated to non-homogeneous regions. BPG builds on the hypothesis that humans are more prone to notice artefacts in complex regions of the image. Following this hypothesis of BPG, the proposed method makes use of object saliency in two ways: 1) {\bf rate optimization}: more bits are allocated to the salient regions; 2) {\bf distortion optimization}: artefacts in salient regions are more heavily penalized. To the best of our knowledge, we are the first to incorporate saliency in learned compression.

The rest of the paper is structured as follows. In Section \ref{sec:percp_sim_metrices}, the compression specific perceptual similarity dataset is presented and various metrics are compared against the human judgements. Section \ref{sec:method} describes the proposed approach which is evaluated in Section \ref{sec:results}. Section \ref{sec:conclusions} concludes the paper.

\section{Perceptual Similarity Metrics}
\label{sec:percp_sim_metrices}

This section investigates various perceptual similarity metrics for both engineered and learned compression methods. We collect a compression specific perceptual similarity dataset and benchmark the existing hand-crafted evaluation metrics PSNR and MS-SSIM, along with a learned metric LPIPS \cite{zhang18}. Following the recent evaluation setup in super-resolution literature \cite{blau2018perception,blau20182018}, we investigate linear combinations of learned and hand-crafted perceptual similarity metrics.

\subsection{Setup for Human Evaluations}
\label{sec:setup_for_human_evaluations}

The setup for collecting this dataset aligns with that of \cite{zhang18} and adapts two alternatives forced choices (2AFC). Annotators are presented with two reconstructed versions of the same image from different compression methods, along with the original image in the middle. They are asked to pick the image which is closer to the original. At high bit rates, the images may be very similar, thus the annotators are provided with a synchronous magnifying glass. They are instructed to scan the images as a whole in cases of uncertainty. The evaluations were hosted on Amazon Mechanical Turk, on average, the annotators spent $56$ seconds on one sample.

The images are obtained using the following image compression methods: Mentzer  ~\etal \cite{mentzer2018conditional}, Patel ~\etal \cite{patel2019deep}, BPG \cite{bpg} and JPEG-2000 \cite{skodras2001jpeg}. A total of $200$ original images are used, comparisons are made at $4$ different bit-rates and all possible combinations of methods are considered, \emph{i.e.}, $6$ combinations for $4$ methods. This results in $4,800$ total samples for perceptual similarity studies. We use $3,840$ samples for training and $960$ held out samples for testing. For each such sample, we obtain $6$ evaluations resulting in a total of $28,800$ annotations.

\subsection{Deep Perceptual Metric}
\label{sec:deep_perceptual_metric}

\begin{figure}
\centering
    \includegraphics[width=0.4\textwidth]{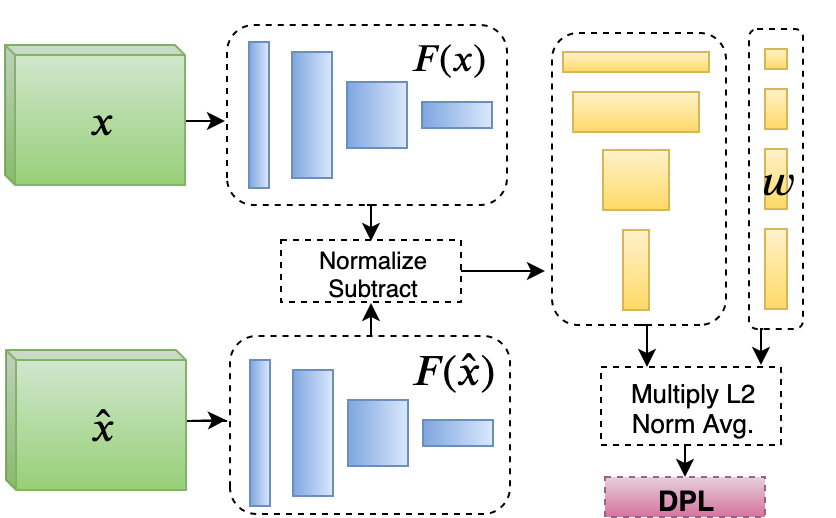}
	\caption{Deep Perceptual Loss: To compute perceptual similarity distance between the original $x$ and the reconstructed $\hat{x}$ images - first compute the deep embeddings $F(x)$ and $F(\hat{x})$, normalize along the channel dimensions, scale each channel vector $w$ (learned on perceptual similarity dataset) and take the $\ell_{2}$ norm. Finally average across spatial dimensions and sum across channels.}
	\label{fig:dpl}
	\vspace{-0.2cm}
\end{figure}

The utility of using deep networks as a deep perceptual similarity metric has been studied by Zhang et al. \cite{zhang18}. It was observed that comparing activations from deep CNNS such as VGG-16 \cite{simonyan2014very} or AlexNet \cite{krizhevsky2012imagenet} acts as a better perceptual similarity metric when compared to MS-SSIM and PSNR. We follow Zhang's approach and make use of activations from five $ReLU$  layers after each $conv$ block in the VGG-16 \cite{simonyan2014very} architecture, with batch normalization.

Feed-forward is performed on VGG-16 for both the original ($\bold{x}$) and the reconstructed image ($\bold{\hat{x}}$). Let $L$ be the set of layers used for loss calculation (five for our setup), a function $F(\bold{x})$ denoting feed-forward on an input image $\bold{x}$. $F(\bold{x})$ and $F(\hat{\bold{x}})$ return two stacks of feature activation's for all $L$ layers. The Deep perceptual loss is then computed as:
\begin{itemize}
\item $F(\bold{x})$ and $F(\bold{\hat{x}})$ are unit-normalized in the channel dimension. Let us call these, $\bold{z^{l}_{x}}, \bold{z^{l}_{\hat{x}}} \in \mathbb{R}^{H_{l} \times W_{l} \times C_{l}}$ where $l \in L$. ($H_{l}, W_{l}$ are the spatial dimensions).
\item $\bold{z^{l}_{x}}, \bold{z^{l}_{\hat{x}}}$ are scaled channel wise by multiplying with the vector $\bold{w^{l}} \in \mathbb{R}^{C_{l}}$ .
\item The $L_{2}$ distance is then computed and an average over spatial dimension is taken. Finally, a channel-wise sum is taken which outputs the deep perceptual loss.
\end{itemize}

Equation \ref{eqn:dpl}  and Figure \ref{fig:dpl} summarize the  Deep perceptual loss computation. Note that the weights in $F$ are learned for image classification on the ImageNet dataset \cite{russakovsky2015imagenet} and are kept fixed. $\bold{w}$ are the linear weights learned on top of $F$ on the perceptual similarity dataset using a ranking loss function \cite{zhang18}. In the next subsection, the \textit{LPIPS} metric is referred to learning $\bold{w}$ on Berkeley-Adobe Perceptual Patch Similarity Dataset \cite{zhang18} and \textit{LPIPS-Comp (Ours)} is referred to the setup when $\bold{w}$ is learned on the compression specific similarity data (Section \ref{sec:setup_for_human_evaluations}). Note that \textit{LPIPS-Comp} is used as the Deep perceptual loss (DPL).

\begin{equation}
\small{DPL(\bold{x},\bold{\hat{x}}) = \sum_{l} \frac{1}{H_{l}W_{l}} \sum_{h,w} || \bold{w_{l}} \odot  (\bold{z^{l}_{\hat{x},h,w}} - \bold{z^{l}_{x,h,w}) ||_{2}^{2}}}
\label{eqn:dpl}
\end{equation}

\subsection{Analysing Metrics for Image Compression}

\begin{figure}
    \centering
    \includegraphics[width=0.5\textwidth]{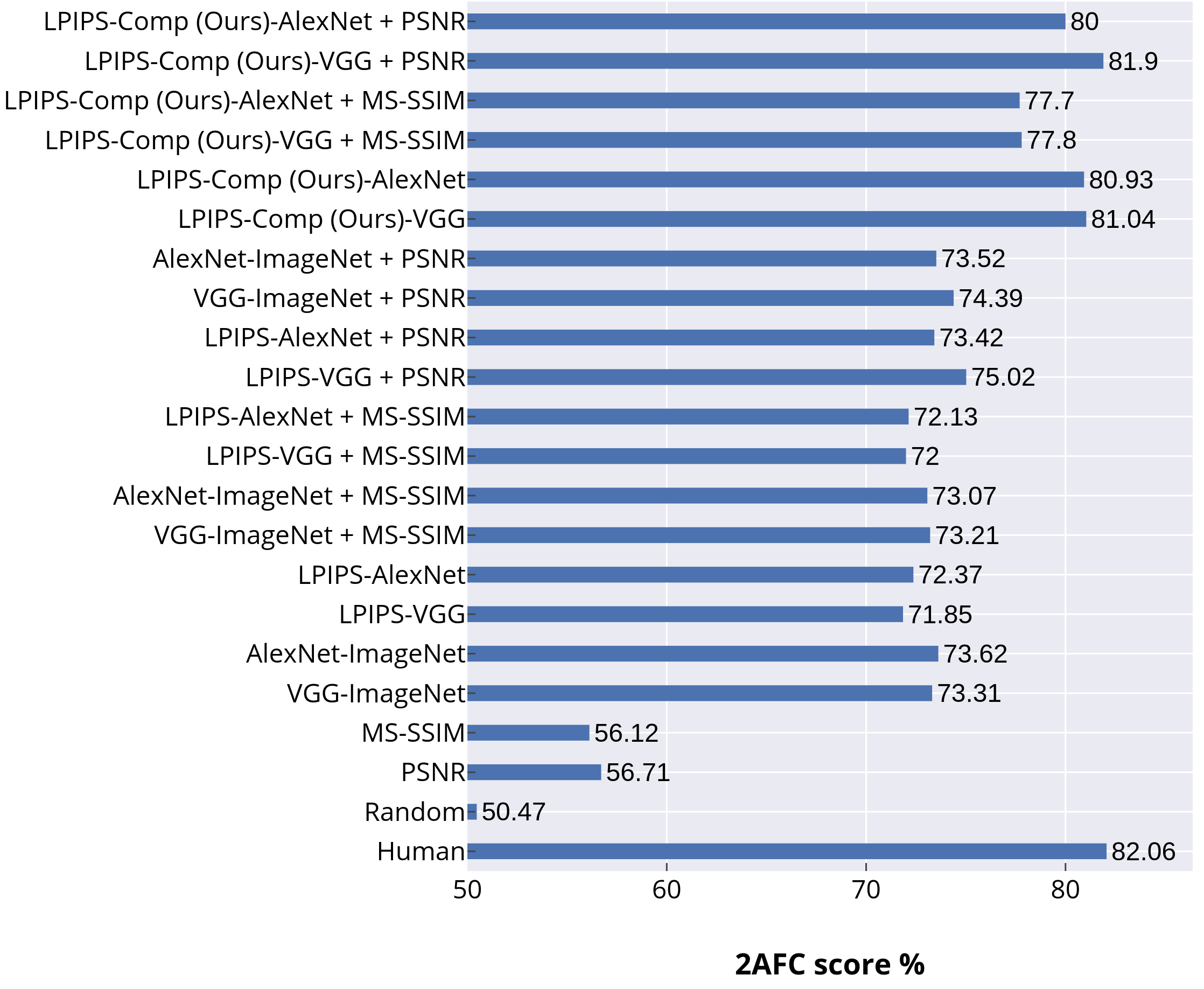}
    \caption{Comparison of various hand-crafted and learnt perceptual similarity metrics on the test set.}
    \label{fig:eval_metrices}
\end{figure}

A comparison of various metrics is provided in Figure \ref{fig:eval_metrices}. The analysis starts by computing the 2AFC score of a human annotator. Then, the performance of two hand-crafted metrics PSNR and MS-SSIM are reported. Note that these are the most popular metrics used in the compression literature to report the state-of-the-art \cite{mentzer2018conditional,lee2018context,balle2018variational,balle2016end,rippel2017real,minnen2018joint}. However, as shown in the figure, the 2AFC scores for them (MS-SSIM and PSNR) are fairly low and do not align well with the human perception of similarity.

A naive similarity metric can be obtained by using AlexNet \cite{krizhevsky2012imagenet} or VGG-16 \cite{simonyan2014very} trained on ImageNet. These features act as a better similarity metric \cite{zhang18} compared to PSNR or MS-SSIM. The features can be adapted better for perceptual similarity by following the framework presented in Section \ref{sec:deep_perceptual_metric}, \emph{i.e.}, linearly re-weighting the channels with the weight vector on a perceptual similarity dataset. When these weights are learned on a generic dataset with a large collection of distortions (\textit{LPIPS} in Figure \ref{fig:eval_metrices}) such as the Berkely-Adobe dataset \cite{zhang18}, the performance is slightly worse compared to directly using the ImageNet model. This indicates a domain gap and establishes that using the similarity data of a different nature can have adverse effects. When the weights are trained on compression specific data (\textit{LPIPS-Comp (Ours)} in  \ref{fig:eval_metrices}), the learned metric aligns much better with the human judgements as can be clearly seen in Figure \ref{fig:eval_metrices}.

Finally, the linear combination of hand-crafted metrics PSNR and MS-SSIM with a learned metric is investigated. Unlike \cite{blau2018perception,blau20182018}, the weights given to each metric are learned on the compression specific similarity dataset. This is achieved by solving a linear optimization problem and employing RANSAC. We refer to the supplementary material for a detailed explanation of learning these weights. It is observed that \textit{LPIPS-Comp (Ours)}, when combined with PSNR almost achieves close to human 2AFC score.

\section{Proposed Method}
\label{sec:method}

\begin{figure*}
\centering
\begin{subfigure}{0.60\textwidth}
    \centering
    \includegraphics[trim={0 0 1.5cm 0},clip,width=\textwidth]{Overall_method.png}
    \caption{\textbf{Overall Method}: In the first stage, the input image is fed to the first encoder and the saliency model, the features are masked using an importance mask and the saliency mask. The masked features are then quantized and fed to the second-stage. Within the second stage the features are fed through another encoder, quantized and compressed independently in a lossless manner using adaptive arithmetic coding. A transformed version of these compressed representations is used to condition the compression (entropy estimation) of the first stage's representation (standard adaptive arithmetic coding is used). Finally the compressed quantized representation are fed to the decoder to generate the reconstructed image.}
    \label{fig:overall_method}
\end{subfigure}
\hspace{0.05\textwidth}
\begin{subfigure}{0.32\textwidth}
    \centering
    \includegraphics[trim={0.5cm 0.5cm 1.5cm 0},clip,width=\textwidth]{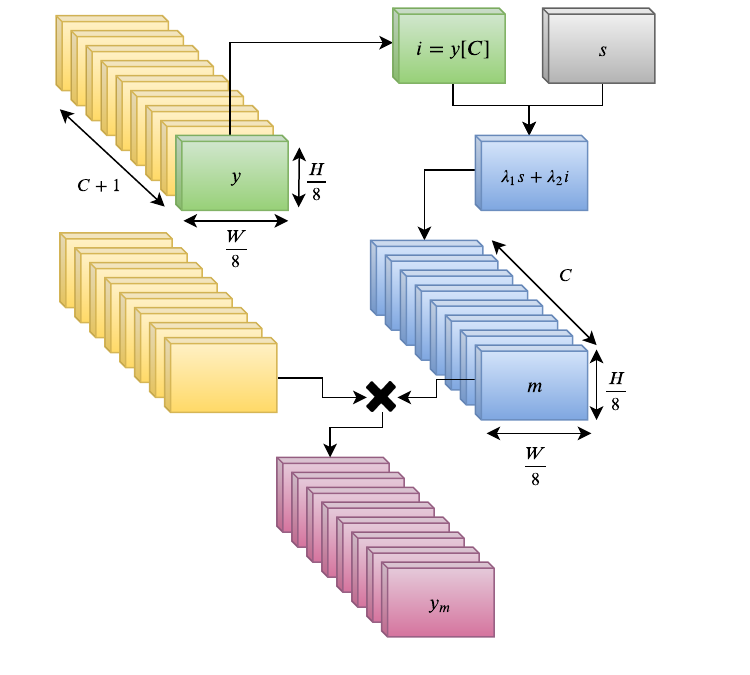}
    \caption{\textbf{Salient Masking}: The last channel of the bottleneck $E_{1}(\bold{x})[C_{1}]$ from the first stage is used as an importance map $\bold{i}$. This importance map is linearly combined with the saliency mask $\bold{s}$ and is expanded to match the dimensions of the bottleneck. Finally the bottleneck is masked using point-wise multiplication.}
    \label{fig:salient_masking}
\end{subfigure}
\caption{\ref{fig:overall_method} provides an overview of our method while \ref{fig:salient_masking} illustrates the proposed saliency driven masking.}
\label{fig:two_figures}
\vspace{-0.5cm}
\end{figure*}

The objective function of any lossy image compression technique is defined by a rate-distortion trade-off: 
\begin{equation}
    \min \sum_{\bold{x} \sim \bold{\chi}}(\alpha Rate(\bold{x}) + \beta Distortion(\bold{x},\bold{\hat{x}}))
    \label{eq:rate_distortion_tradeoff}
\end{equation}
where $\bold{x}$ is the image to be compressed drawn from a collection of images $\bold{\chi}$, $\bold{\hat{x}}$ is the reconstructed image, $Rate(\bold{x})$ is the storage requirement for the image and $Distortion(\bold{x},\bold{\hat{x}})$ is a measure of distortion between the original and the reconstructed images.

As shown in Figure \ref{fig:overall_method}, our method consists of two encoders, two decoders, two quantization stages and two auto-regressive models for entropy estimation. All models are trained jointly and in an end-to-end fashion. The first encoder takes the image as input and outputs latent representation $\bold{y}=E_{1}(\bold{x}): \mathbb{R}^{W \times H \times 3} \xrightarrow{} \mathbb{R}^{\frac{W}{8} \times \frac{H}{8} \times C_{1}+1}$ (Section \ref{sec:encoder_decoder}). Note that the number of channels in the bottleneck, $C_{1}$, is one of the hyper-parameters to train the models to obtain different bits-per-pixel values. A pre-trained network outputs the object saliency for the input image $\bold{s}=S(\bold{x}): \mathbb{R}^{W \times H \times 3} \xrightarrow{} \mathbb{Z}_{2}^{\frac{W}{8} \times \frac{H}{8} \times 1}$ (Section \ref{sec:saliency_matters}). The latent representations are first masked by saliency driven priors $\bold{y_{m}}=m_{1}(\bold{y},\bold{s})$ (Section \ref{sec:saliency_matters}) and then quantized $\bold{\Tilde{y}}=Q_{1}(\bold{y_{m}}): \mathbb{R} \xrightarrow{} \{c_{(1,1)}, ... , c_{(1,L)}\}$ (Section \ref{sec:quantization}) and fed to stage-two. Within stage two, the second encoder outputs the latent representations $\bold{z}=E_{2}(\bold{\Tilde{y}}): \mathbb{R}^{\frac{W}{8} \times \frac{H}{8} \times C_{1}} \xrightarrow{} \mathbb{R}^{\frac{W}{32} \times \frac{H}{32} \times C_{2}+1}$ which are also masked $\bold{z_{m}}=m_{2}(\bold{z})$ (independent of saliency) and quantized with different centers $\bold{\Tilde{z}}=Q_{2}(\bold{z_{m}}): \mathbb{R} \xrightarrow{} \{c_{(2,1)}, ... , c_{(2,L)}\}$.

An auto-regressive image compression model operating on a quantized latent representation \cite{mentzer2018conditional} factorizes the discrete representation using a basic chain rule \cite{kolesnikov2017pixelcnn}: 
\begin{equation}
    P(\bold{\Tilde{y}}) = \prod_{i=1}^{N}p(\Tilde{y}_{i}|\Tilde{y}_{i-1}, ..., \Tilde{y}_{1})
\end{equation}

Our idea is to jointly learn an extra set of auxiliary representations $\bold{\Tilde{z}}$ to factorize joint distribution using:
\begin{equation}
    P(\bold{\Tilde{y}}, \bold{\Tilde{z}}) = P_{\Theta}(\bold{\Tilde{y}}| \bold{\Tilde{z}})P_{\Phi}(\bold{\Tilde{z}})
\end{equation}

Here $\Theta$ and $\Phi$ are the parameters of two 3D Pixel-CNN models where $P(\bold{\Tilde{z}})$ is the second stage which is decoded first during the decompression stage. Thus for the first stage, quantized representations $\bold{\Tilde{y}}$ are encoded and decoded by assuming that the $\bold{\Tilde{z}}$ is available. In the subsequent sections, each component of the proposed method is described. 

\subsection{Encoder-Decoder}
\label{sec:encoder_decoder}

The method consists of two encoders and two decoders. The first encoder is a fifteen residual blocks \cite{he2016deep} fully-convolutional network with three non-local/self-attention layers \cite{wang2018non}. The first encoder involves down-sampling of the input image $\bold{x}$ by a factor of $8$. The second encoder takes the quantized representations from the first stage $\bold{\Tilde{y}}$ as input, feed-forwards through five residual blocks \cite{he2016deep}, a non-local layer \cite{wang2018non} and involves a further down-sampling by a factor of $4$. $\bold{\Tilde{z}}$ is $\frac{W}{32}\times \frac{H}{32}$ and fairly small compared to the input $\bold{x}$ of $W\times H$. Thus, the number of bits required to store the second stage bit-string is very low (roughly $5\%$ of the total storage). The decoders corresponding to these two encoders are their mirror.

The number of channels in the bottlenecks $\bold{\Tilde{y}}$ or $\bold{\Tilde{z}}$ is a hyper-parameter used to control the bits-per-pixel. In practice, the number of channels is kept the same, for both of these bottlenecks, \emph{i.e.}, $C_{1}=C_{2}$. Both bottlenecks have an extra channel for a hybrid of saliency mask and an importance map (Section \ref{sec:saliency_matters}).

\subsection{Quantization}
\label{sec:quantization}

Quantization is a non-differentiable function, with gradients being either zero or infinite, thus any deep learning model with quantization cannot be trained using backpropagation \cite{rumelhart1986learning}. Thus, soft vector quantization \cite{agustsson2017soft} is adapted in our model. More specifically, given a set of cluster centers $C_{1}=\{c_{1},...,c_{L_{1}}\}$ the feed-forward is determined by:

\begin{equation}
    \Tilde{y}_{i} = Q_{C_{1}}(y) = arg min_{j} || y_{i} - c_{j} ||
\end{equation}

during backpropagation, a soft cluster assignment is used:

\begin{equation}
    \hat{y}_{i} = \sum_{j=1}^{L_{1}} \frac{exp(-\sigma ||y_{i}-c_{j}||)}{\sum_{l=1}^{l=L_{1}}exp(-\sigma ||y_{i}-c_{j}||)} 
    \label{eq:dpl}
\end{equation}

Note that the quantization process is the same for both stages, but with different sets of centres.

\subsection{Hierarchical Auto-Regressive Model}

\paragraph{First Stage.}
The representations of the first stage are encoded and decoded by conditioning on the second stage and may be fully factorized as:

\begin{equation}
P(\bold{\Tilde{y}} | \bold{\Tilde{z}}) = \prod_{i=1}^{i=N} P(y_{i} | y_{i-1},...,y_{1},D_{2}(\Tilde{z}))    
\end{equation}

The quantized representations of the first stage are losslessly compressed using standard arithmetic coding where the conditional probabilities are estimated by a 3D pixel-CNN \cite{kolesnikov2017pixelcnn} which is conditioned on extra auxiliary representations $D_{2}(\Tilde{z})$. The 3D pixel-CNN is trained with cross-entropy minimization for a correct quantized centre assignment:
\begin{equation}
    P_{i,l}(\bold{\Tilde{y}}) = p_{\Theta}(\Tilde{y}_{i}=c_{l}|\Tilde{y}_{i-1},...,\Tilde{y}_{1},D_{2}(\bold{\Tilde{z}}))
\end{equation}
Thus the total entropy for the bottleneck is estimated using cross-entropy as:
\begin{equation}
    CE = H_{1}(\bold{\Tilde{y}}|\bold{\Tilde{z}}) = \mathbb{E}_{\bold{\Tilde{y}}\sim P(\bold{\Tilde{y}}|\bold{\Tilde{z}})}[\sum_{i=1}^{i=N}-log(P_{i,l}(\Tilde{y}_{i}))]
    \label{eq:first_stage_entropy}
\end{equation}

\vspace{-0.3cm}
\paragraph{Second Stage.}
The representations of the second stage are encoded independently and the distribution is factorized as a product of conditionals:
\begin{equation}
    P(\bold{\Tilde{z}}) = \prod_{i=1}^{i=M} P(\Tilde{z}_{i} | \Tilde{z}_{i-1},...,\Tilde{z}_{1})
\end{equation}

The second stage uses a separate 3D pixel-CNN \cite{kolesnikov2017pixelcnn}, which is trained by minimizing:
\begin{equation}
    CE = H_{2}(\bold{\Tilde{z}}) = \mathbb{E}_{\bold{\Tilde{z}}\sim P(\bold{\Tilde{z}})}[\sum_{j=1}^{j=M}-log(P_{j,l}(\Tilde{z}_{j})]
\end{equation}

The objective of the second stage is to learn the auxiliary features which help in compressing the first stage representations. Thus the gradients from Equation \ref{eq:first_stage_entropy} are propagated to the second stage along with additional gradients from a reconstruction loss $mse(\bold{\Tilde{y}}, D_{2}(\bold{\Tilde{z}}))$.

\vspace{-0.3cm}
\paragraph{Joint Optimization.} The overall rate of optimization $Rate(\bold{x})$ incorporates the masks from both stages, that is $\bold{m_{1}}$ and $\bold{m_{2}}$ as the weights to the cross-entropy computation for a given index in the bottleneck. The overall entropy is thus given by:
\begin{equation}
    Rate(\bold{x}) = H = \bold{m_{1}}H_{1}(\bold{\Tilde{y}}|\bold{\Tilde{z}}) + \bold{m_{2}}H_{2}(\bold{\Tilde{z}})
    \label{eq:rate_optimization}
\end{equation}

\subsection{Incorporating Object Saliency}
\label{sec:saliency_matters}

The saliency mask $\bold{s}$ such that $s_{i} \in \{0,1\}$ is predicted by an off-the-shelf object saliency model \cite{hou2017deeply}, which was trained on MSRA10K data \cite{cheng2014global}. It is used in our compression model in two ways. Firstly, to mask quantized representations of the first stage, that is, more bits are allocated to the salient regions. Secondly, during the computation of distortion loss, to give higher weight to the reconstruction of the salient regions.

\vspace{-0.3cm}
\paragraph{Salient Masking.} The generated saliency mask is combined with an importance mask which helps in navigating the bit-rate convergence to a certain target value \cite{mentzer2018conditional}. Similar to \cite{mentzer2018conditional}, the last channel of the bottleneck is treated as the importance mask $\bold{i}=E_{1}(\bold{x})[C_{1}]$. This importance mask is linearly combined with the saliency mask $\bold{s}$ to make the compression driven by saliency. As illustrated in Figure  \ref{fig:salient_masking}, the final mask used is given by $\bold{m}_{1} = \lambda_{1}\bold{s} + \lambda_{2}\bold{i}$. In practice, $\lambda_{1}=\lambda_{2}=1$, this way the model is able to incorporate saliency while at the same time it is able to converge to a specified target bit-rate value.

This two-dimensional mask is expanded to match the dimensionality of the bottleneck \cite{mentzer2018conditional}. Finally, the bottleneck is masked by a pointwise multiplication with the binarization of $\bold{m}_{1}$ as $\bold{y}_{m} = \bold{y} \odot \lceil{\bold{m}_{1}} \rceil$.


\vspace{-0.3cm}
\paragraph{Weighted Distortion Losses.} We hypothesise that humans, in general, pay more attention to salient regions in the image. Thus, during training, a higher priority is given to the reconstruction of salient regions. This is achieved by decomposing the original and reconstructed images into salient and non-salient parts. Here distortion loss is computed on both separately and then linearly combined as:
\begin{equation}
    w_{1}D(\bold{x}\odot \bold{s}, \bold{\hat{x}}\odot \bold{s}) +  w_{2}D(\bold{x}\odot (\bold{1-s}), \bold{\hat{x}}\odot (\bold{1-s}))
    \label{eq:salient_distortion}
\end{equation}

Where $w_{1}>w_{2}$, in practice, $w_{1}=0.75$ and $w_{2}=0.25$.  Refer to the supplementary material for an illustration.

\subsection{Model Optimization}
\label{sec:overall_optimization}

The overall optimization is a rate-distortion trade-off (Equation \ref{eq:rate_distortion_tradeoff}). The rate is determined by Equation \ref{eq:rate_optimization}, the distortion is saliency driven and is governed by Equation \ref{eq:salient_distortion} where the distortion function $D$  is a linear combination of the Deep perceptual loss DPL (Equation \ref{eqn:dpl}) (LPIPS-Comp) and the mean-squared error between the original and the reconstructed images.

\vspace{-0.3cm}
\paragraph{Training Details.}
\label{sec:training_details}
Adam optimizer \cite{kingma2014adam} is used, with an initial learning rate of $4 \times 10^{-3}$ and a batch-size of $30$. The learning rate is decayed by a factor of $10$ in every two epochs (step-decay). Further, similar to \cite{mentzer2018conditional}, the rate term is clipped as $max(t, \beta R)$, to make the model converge to a certain bit-rate $t$. The training is done on the training set of ImageNet dataset the from Large Scale Visual Recognition Challenge 2012 (ILSVRC2012) \cite{russakovsky2015imagenet}, with the mentioned setup, convergence was observed in six epochs.

By varying the model hyper-parameters such as the number of channels in the bottlenecks (that is $C_{1}$ and $C_{2}$), the weight for distortion loss ($\alpha$), the target bit-rate ($t$), multiple models were obtained in the bit-per-pixel range of $0.15$ to $1.0$. Similarly, the models for \cite{mentzer2018conditional,balle2016end} were reproduced at different bits-per-pixel values. Note that, in the case of \cite{balle2016end}, we used an MS-SSIM loss instead of MSE loss as was done in the original paper but this does not change the general conclusions of the paper. For Lee \etal \cite{lee2018context}, the authors provided us with the images from the Kodak dataset for human evaluations.

\section{Results}
\label{sec:results}
The results of image compression are shown using human evaluations. Another way to compare compression methods is to judge how well the reconstructed images do on a computer vision task. The model for this task is pre-trained on uncompressed images, specifically, object detection and segmentation are investigated. Each lossy compression method creates a certain kind of artefacts that impact the task. For example, Dwibedi ~\etal \cite{dwibedi2017cut} show that for object detectors such as Faster-RCNN \cite{fasterrcnn} region-based consistency is important and pixel-level artefacts can affect the performance. Thus a compression method which distorts the region based consistency will perform poorly for object detection. We view this as another evaluation of a compression technique. We demonstrate that on all three metrics our approach outperforms competing methods. Section \ref{sec:percp_sim_metrices} shows that the widely used metrics PSNR and MS-SSIM are inadequate for judging different compression methods. However, for completeness, results on PSNR and MS-SSIM are also reported.

\vspace{-0.3cm}
\paragraph{Comparison with other compression methods.} We compare the proposed method with the state-of-the-art image compression models from Lee~\etal \cite{lee2018context} based on variational autoencoders. We also compare to a single-level 
auto-regressive compression method Mentzer et al \cite{mentzer2018conditional} and two engineered methods BPG \cite{bpg} and JPEG-2000 \cite{skodras2001jpeg}. We use the Kakadu\footnote{http://kakadusoftware.com/} implementation for JPEG-2000 and use BPG in the 4:4:4 chroma format following \cite{mentzer2018conditional,rippel2017real}.

\begin{figure}
    \centering
    \includegraphics[width=\linewidth]{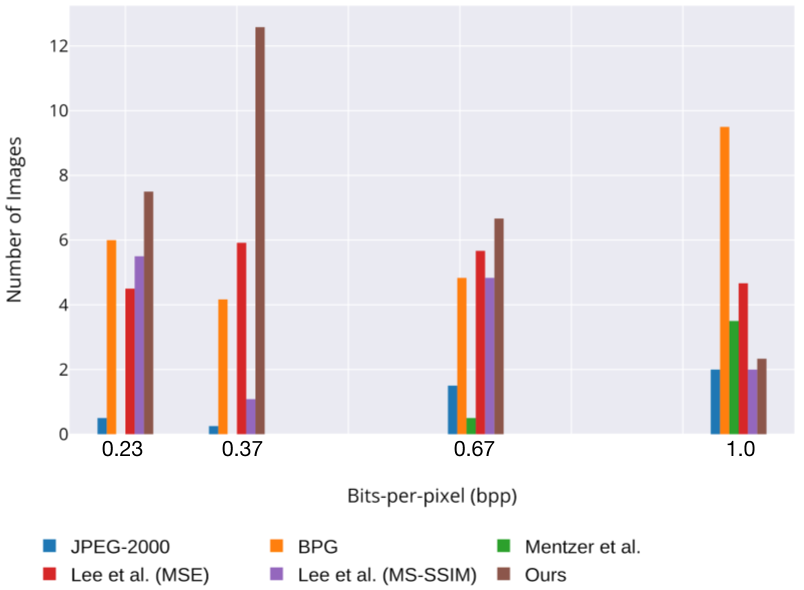}
    \caption{Human evaluations on the Kodak dataset. The y-axis shows the number of images for which a given method performs best. The x-axis shows the BPP values at which the comparisons were performed.}
        \label{fig:kodak_human_evals}
        \vspace{-0.2cm}
\end{figure}

\vspace{-0.3cm}
\paragraph{Quantitative Human Evaluations.} An extensive human evaluation study is performed using five compression approaches, across four different bits-per-pixel values (0.23, 0.37. 0.67, 1.0) on the Kodak dataset \cite{kodak}. For the human evaluation, we follow the setup described in Section \ref{sec:setup_for_human_evaluations}. The comparison is made in a pair-wise manner for all $15$ possible combinations of the six methods ($^{6}C_{2}$). For each such pair-wise comparison, we obtain five evaluations and determine the better performing method. Across different pair-wise comparisons, the method which wins the most number of times performs best for the given image at a bit-rate. Thus, at each bit-rate, we count the number of images for which a method performs best among the set of competing methods. Figure \ref{fig:kodak_human_evals} shows that our method is best according to the human evaluation across three bit-rate values (0.23, 0.37. 0.67). At a relatively higher BPP of 1.0 BPG out-performs all the other methods.

\begin{figure*}
    \centering
    \includegraphics[width=\textwidth]{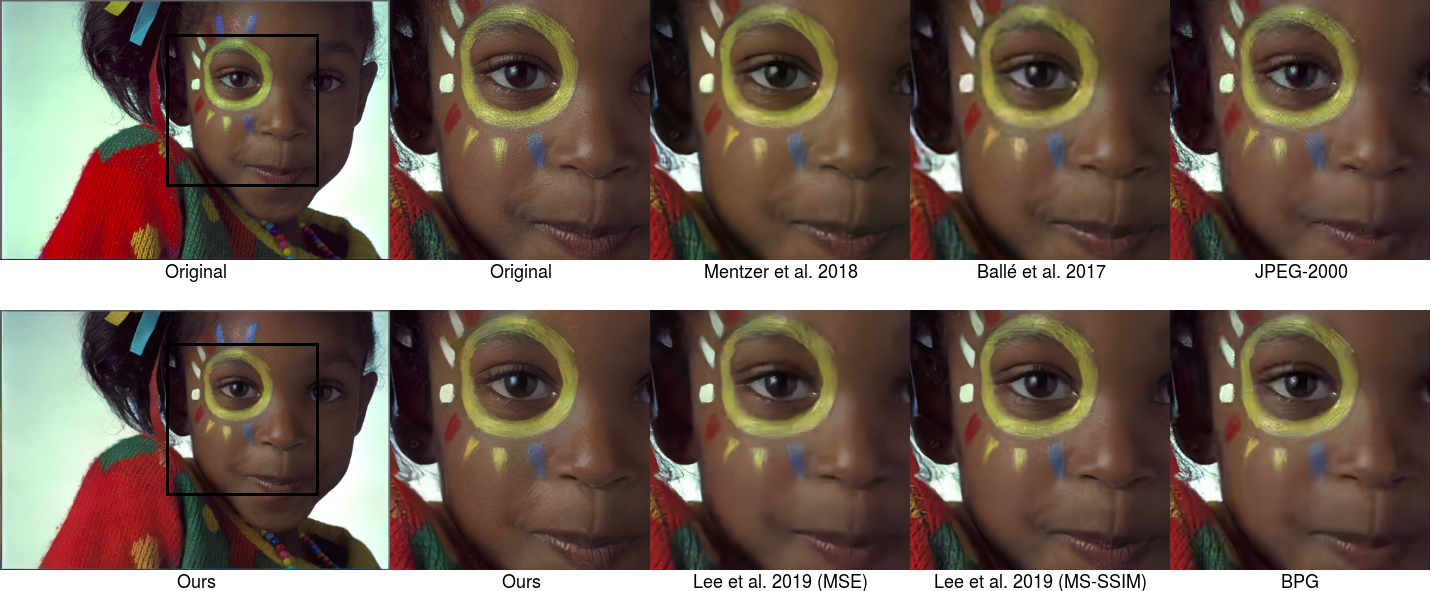}
    \caption{A qualitative example from the Kodak dataset \cite{kodak} at \textbf{0.23} BPP. Notice our method better captures the fine-grained details (lines above the lip, yellow circle around the eye) better than other approaches. }
    \label{fig:quanlitative_results}
\end{figure*}

\vspace{-0.3cm}
\paragraph{Qualitative Comparison.} Please see Figure \ref{fig:quanlitative_results} where a comparison from Kodak dataset is shown. Notice that, our method preserves the fine-grained details of the face better than the other methods (see above the lip, paint patterns around the eye). For more such examples we refer the reader to the supplementary material.

\begin{table}
\centering
\small
\begin{tabular}{l | c | c | c | c}
\toprule
\textbf{Method} & \textbf{0.23} & \textbf{0.37} & \textbf{0.67} & \textbf{1.0}\\
\midrule
JPEG-2000 \cite{skodras2001jpeg} & 23.2 & 29.1 & 34.4 & 36.8\\
BPG \cite{bpg} & 25.2 & \underline{32.5} & 35.4 & \underline{37.7}\\
Mentzer ~\etal \cite{mentzer2018conditional} & 25.5 & 30.2 & 34.5 &  36.6\\
Lee ~\etal \cite{lee2018context} (MSE) & \underline{28.3} & - & \underline{36.2} & 37.6\\
Lee ~\etal \cite{lee2018context} (MS-SSIM) & 27.2 & \underline{32.5} & - & 37.6 \\
\midrule
Ours (MSE + DPL) & \textbf{29.3} & \textbf{33.7} & \textbf{36.6} & \textbf{37.9}\\
\bottomrule
\end{tabular}
\vspace{-0.2cm}
\caption{Object Detection on MS-COCO 2017 \cite{lin2014microsoft} validation set using Faster-RCNN \cite{fasterrcnn}. The performance is reported using AP@[.5:.95], that is an average over different scales of IoU. Note that the performance on the original (uncompressed) images is $40.1\%$.}
\label{table:object_detection}
\end{table}

\begin{table}
\centering
\small
\begin{tabular}{l | c | c | c | c}
\toprule
\textbf{Method} & \textbf{0.23} & \textbf{0.37} & \textbf{0.67} & \textbf{1.0}\\
\midrule
JPEG-2000 \cite{skodras2001jpeg} & 20.2 & 25.4 &  30.1 & \underline{32.2} \\
BPG \cite{bpg} & 22.0 & 28.5 & 30.8 & \underline{32.2}\\
Mentzer ~\etal \cite{mentzer2018conditional} & 9.3 & 10.5 & 11.9 & 22.0 \\
Lee ~\etal \cite{lee2018context} (MSE) & \underline{25.4} & - &  \underline{32.2} & \textbf{33.2}\\
Lee ~\etal \cite{lee2018context} (MS-SSIM) & 25.1 & \underline{28.9} & - & \textbf{33.2} \\
\midrule
Ours (MSE + DPL) & \textbf{26.1} & \textbf{30.0} & \textbf{32.3} & \textbf{33.2}\\
\bottomrule
\end{tabular}
\vspace{-0.2cm}
\caption{Instance segmentation on MS-COCO 2017 \cite{lin2014microsoft} validation set using Mask-RCNN \cite{he2017mask}. The performance is reported using an average over multiple IoU values. Note that the performance on the original (uncompressed) images is $35.2\%$.}
\label{table:instance_segmentation}
\vspace{-0.5cm}
\end{table}

\vspace{-0.3cm}
\paragraph{Object Detection.} A pre-trained Faster-RCNN \cite{fasterrcnn} model with a $ResNet-101$\cite{he2016deep} based backbone is used. With the original MS-COCO images, this model attains a performance of $40.1\%$ AP. For each compression method, we compress and reconstruct the image at four different bit-rate values: $0.23, 0.37, 0.67, 1.0$ (same values as used for human evaluation) and the reconstructed images are evaluated for object detection. The performance of competing compression methods are reported in Table \ref{table:object_detection}. It can be seen that the proposed method outperforms the competing methods at all bit-rates.

\vspace{-0.3cm}
\paragraph{Instance segmentation.} Mask-RCNN \cite{he2017mask} with a \textit{ResNet-101} backbone is used for the task. The performance of different compression techniques is reported in Table \ref{table:instance_segmentation}. It is observed that while Mentzer ~\etal \cite{mentzer2018conditional} perform comparable to engineered methods on object detection, it performs far worse on Instance segmentation. It can be seen in Table \ref{table:instance_segmentation}, that our method outperforms the competing methods at lower bit-rates while \cite{lee2018context} performs identically at $1.0$ bits-per-pixel.

\begin{figure*}
    \centering
    \includegraphics[width=\textwidth]{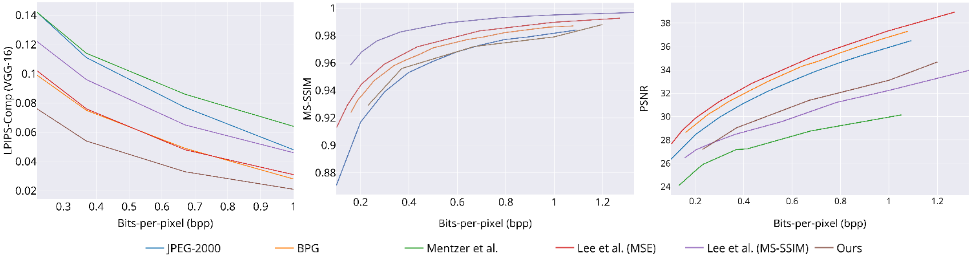}
    \caption{Different metrics on the Kodak dataset \cite{kodak}. \textit{\textbf{Left}}: (Perceptual Similarity) LPIPS-Comp (VGG-16) (Section \ref{sec:deep_perceptual_metric}) vs bits-per-pixel (BPP) (Lower is better). \textit{\textbf{Middle}}: MS-SSIM vs bits-per-pixel (BPP). (Higher is better). \textit{\textbf{Right}}: PSNR vs bits-per-pixel (BPP). (Higher is better). Best viewed in color.}
    \label{fig:distortion_evals}
    \vspace{-0.4cm}
\end{figure*}

\vspace{-0.3cm}
\paragraph{PSNR / MS-SSIM / LPIPS-Comp.} For completeness, the performance of these models are shown on standard PSNR and MS-SSIM metrics. Figure \ref{fig:distortion_evals} (right) for PSNR, Figure \ref{fig:distortion_evals} (middle) for MS-SSIM and Figure \ref{fig:distortion_evals} (left) for LPIPS-Comp (Section \ref{sec:deep_perceptual_metric}). 

\section{Conclusions}
\label{sec:conclusions}
An approach is proposed for deep image compression. The method trains and evaluates using a deep perceptual metric. It uses a hierarchical auto-regressive framework and takes object saliency into account. On human evaluations, the model performs better than the state-of-the-art on low bit rates. Images obtained from our model provide the best object detector and image segmentation results when compared to the other image compression schemes.

\section*{Acknowledgement}

The authors thank Joel Chan and Peter Hallinan for helping with the human evaluations.

{\small
\bibliographystyle{ieee_fullname}
\bibliography{egbib}
}

\end{document}